# Properties of Fr- like $Th^{3+}$ from spectroscopy of high-L Rydberg levels of $Th^{2+}$

Julie A. Keele, M.E. Hanni, Shannon L. Woods and S.R. Lundeen

*Department of Physics, Colorado State University, Fort Collins, Colorado 80523, USA*

C.W. Fehrenbach

*J.R. Macdonald Laboratory, Kansas State University, Manhattan, Kansas 66506, USA*

Binding energies of high-$L$ Rydberg states ($L \geq 7$) of $Th^{2+}$ with $n$=27-29 were studied using the Resonant Excitation Stark Ionization Spectroscopy (RESIS) method. The core of the $Th^{2+}$ Rydberg ion is the Fr-like ion $Th^{3+}$ whose ground state is a $5\ ^2F_{5/2}$ level. The large core angular momentum results in a complex Rydberg fine structure pattern consisting of six levels for each value of $L$ that is only partially resolved in the RESIS excitation spectrum. The pattern is further complicated, especially for the relatively low $L$ levels, by strong non-adiabatic effects due to the low-lying $6d$ levels. Analysis of the observed RESIS spectra leads to determination of five properties of the $Th^{3+}$ ion: its electric quadrupole moment, Q = 0.54(4), its adiabatic scalar and tensor dipole polarizabilities, $\alpha_{d,0}$ = 15.42(17) and $\alpha_{d,2}$ = -3.6(1.3), and the dipole matrix elements connecting the ground $5^2F_{5/2}$ level to the low lying $6^2D_{3/2}$ and $6^2D_{5/2}$ levels, $|<5^2F_{5/2}||D||6^2D_{3/2}>|$ = 1.435(10) and $|<5^2F_{5/2}||D||6^2D_{5/2}>|$ = 0.414(24). All are in atomic units. These are compared with theoretical calculations.

## I. INTRODUCTION

High-$L$ non-penetrating Rydberg electrons act as sensitive probes of the properties of the core ion, such as permanent moments and polarizabilities that control its long-range interactions. The hydrogenic degeneracy of Rydberg levels of common $n$, present when the only core property is net charge, is broken by the presence of additional long-range interactions. These produce a fine structure pattern whose shape and scale can be related to the core properties. The high-$L$ Rydberg eigenstates are characterized by the total angular momentum exclusive of Rydberg spin:

$$\vec{K} = \vec{L} + \vec{J}_c, \qquad (1)$$

and levels are labeled by their $n$, $L$, $K$ values. In general, the high-$L$ binding energies are only slightly different from hydrogenic, and for that reason special experimental techniques are needed to observe them. One method that has been used to study many such fine structure patterns is the Resonant Excitation Stark Ionization Spectroscopy (RESIS) technique [1]. This method was recently used to study Rydberg levels of Ni whose complexity is comparable to the $Th^{2+}$ levels studied here [2].

The $Th^{3+}$ ion is Fr-like, since it has a single valence electron outside a closed shell Rn-like core, but unlike neutral Fr its ground state is not a $7\,^2S_{1/2}$ state. Instead the increased nuclear charge leads to a $5\,^2F_{5/2}$ ground state [3, 4]. Some limited optical spectroscopy has been reported for $Th^{3+}$ [5], determining the relative positions of the lowest 24 levels, but no other $Th^{3+}$ properties have ever been measured. Measurements of properties such as polarizabilities and permanent moments provide a valuable test of the very challenging *a-priori* theoretical calculations used to predict the behavior of this ion. Increased confidence in these calculations is important for applications where the isolated ion is used directly [6], where it interacts with other atoms or molecules [7], and where it is imbedded in chemical compounds or solids [8].

Section II of this paper discusses the experimental technique used in this study and reports the RESIS spectra observed. Section III describes the analysis of the spectra with particular attention to the strong non-adiabatic effects that are present. Then Section IV compares the experimental measured properties with theoretical calculations.

## II. EXPERIMENT

The measurements reported here were carried out with a method and apparatus used recently for studies of $Pb^{2+}$, $Pb^{4+}$ [9], and $Th^{4+}$ [10]. The apparatus is illustrated schematically in Fig. 1. For this study, a beam of 75 keV $Th^{3+}$ is produced by sputtering Th metal in a 14-GHz permanent magnet electron cyclotron resonance ion source. The $Th^{3+}$ beam is then charge and mass selected using a 20° magnet and focused by a set of electrostatic quadrupole doublets. The $Th^{3+}$ beam then intersects a Rb Rydberg target where a small fraction of the beam (~ 1-3 %) captures a single Rydberg electron, becoming a $Th^{2+}$ Rydberg beam. The $Th^{2+}$ Rydberg beam is charge selected by a 15° magnet and focused by an electrostatic lens. The electric field in this lens also serves to ionize very weakly bound Rydberg states of $Th^{2+}$ that would otherwise be ionized in the Rydberg detector, producing an undesirable background to the RESIS signal. A $CO_2$ laser beam then intersects the Rydberg beam and excites transitions from one $n$ level to a much higher $n$ level. Several RESIS transitions were studied for this work, exciting $n$ = 27, 28, and 29 levels. Table I shows the transitions studied and the $CO_2$ laser line used for each. The $CO_2$ laser is Doppler tuned by varying the angle of intersection between the Rydberg beam and the fixed frequency $CO_2$ laser beam. The conversion of the intersection angle into the Doppler tuned frequency is discussed in reference [10]. After passing through the $CO_2$ laser beam, the Rydberg ion beam enters the Stark Ionization detector where it encounters a sequence of electrodes that are adjusted to fully ionize and decelerate ions excited by the $CO_2$ laser in order to distinguish them from any $Th^{3+}$ ions that may be present in the beam due to autoionization or collisional ionization. The resulting voltage labeled $Th^{3+}$ ions are deflected into a channel electron multiplier and the current synchronous with chopping of the $CO_2$ laser is measured with a lock-in amplifier.

By scanning the frequency of the Doppler-tuned $CO_2$ laser through a range of frequencies near the hydrogenic transition frequency it is possible resolve the fine structure of the lower n level. Three RESIS spectra were observed in this study, exciting *n*=27, 28, and 29. These are illustrated in Fig. 2. To facilitate comparison between the three spectra, the transition frequencies are rescaled as quasi quantum defects, according to:

$$\delta \equiv \frac{n^3}{2Z^2} \frac{h(\upsilon - \upsilon_0)}{Ryd_M} \qquad (2)$$

where $\upsilon$ is the transition frequency, $\upsilon_0$ is the non-relativistic hydrogenic frequency for the transition, and *n* is the principal quantum number of the level being excited. The non-relativistic hydrogenic frequency, $\upsilon_0$, is defined by:

$$\upsilon_0 = Z^2 Ryd_M \left( \frac{1}{n^2} - \frac{1}{n'^2} \right) \qquad (3)$$

where the Z is the charge of the core and $Ryd_M$ is the mass corrected Rydberg constant. The quasi quantum defect ignores the contribution to the excitation energy from the fine structure in the upper level, which can be significant, but it nevertheless accounts for most of the difference in magnitude of the fine structure in the three levels excited. The range of $\delta$ in Fig. 2 corresponds to a frequency range of -2 to +14 GHz from hydrogenic. The scan range was limited by the speed of the $Th^{3+}$ beam and physical limitations of the Doppler-tuning stages.

One obvious feature in all three spectra of Fig. 2 is the large peak near $\delta = 0$. This is due to the highest $L$ levels being excited, whose quantum defects are very small. The splitting of this peak into two resolved features is a signature of the permanent quadrupole moment of the $Th^{3+}$ ion, which is the dominant feature of the structure at extremely high-$L$. The shape of this feature can give rough estimates of the quadrupole moment and scalar dipole polarizability of the $Th^{3+}$ ion. More detailed information must rely on the resolved features at larger $\delta$, which are shown in Fig. 2 where the vertical scale is magnified by a factor of 20.

## III. ANALYISIS AND RESULTS

### A. The Adiabatic Model

Previous studies of Rydberg systems with anisotropic cores include Ne and Ar ($J_c$=3/2) [12, 13] and Ni ($J_c$=5/2) [2]. Each of these systems were adequately described by an effective potential derived from an adiabatic expansion of the second-order perturbation energies. This approach of finding the effective potential was first developed by Drachman for helium Rydberg levels [14]. The derivation of the effective potential having to do with systems of higher $J_c$ is discussed in references [1] and [15]. The most significant terms in such a potential are shown in Eq. 4. Three core properties, the quadrupole moment, Q, and the adiabatic scalar and tensor dipole polarizabilities, $\alpha_{d,0}$ and $\alpha_{d,2}$, determine the gross fine structure of any system described by such a potential.

$$V_{eff} = -\left[\frac{e^2 \alpha_{d,0}}{2r^4}\right] - \left[\left[\frac{eQ}{r^3} + \frac{\alpha_{d,2}}{2r^4}\right] \times \frac{X_c^{[2]}(J_c) \cdot C^{[2]}(\Omega_r)}{\begin{pmatrix} 5/2 & 2 & 5/2 \\ -5/2 & 0 & 5/2 \end{pmatrix}}\right] \quad (4)$$

A consequence of such a description is that the fine structure pattern would show only minor variations in scale as the principal quantum number changed, showing roughly constant quantum defects. Close examination of the three spectra shown in Fig. 2 shows that this is not the case in the Th$^{2+}$ Rydberg levels studied here. In the relatively well-resolved portions of the three spectra, there are clear variations in the appearance of the three spectra. This is an indication that the adiabatic polarization potential cannot describe the structure. The failure of the adiabatic polarization potential was expected because it is known that the *6d* levels of Th$^{3+}$ lie very close to the ground *5$^2$F$_{5/2}$* level. The excitation energy of the *6$^2$D$_{3/2}$* level is 9193cm$^{-1}$, and that of the *6$^2$D$_{5/2}$* level is 14486cm$^{-1}$ [5]. Much larger excitation energy would be required to assure the adequacy of the adiabatic expansion. A similar circumstance was encountered in studies of Rydberg levels of barium by Gallagher [16] and Snow [17]. In that system, the low-lying *5d* levels produced strong non-adiabatic effects in the quadrupole polarization of the $^2S_{1/2}$ ground state of Ba$^+$.

### B. Non-Adiabatic Effects

To appreciate the limitations of the adiabatic model, consider the portion of the second-order dipole perturbation energy of a Rydberg state with quantum number *n, L, K* due to intermediate states where the Th$^{3+}$ core ion is in the 6$^2$D$_{3/2}$ level.

$$E^{[2]}_{n,L,K}(6^2 D_{3/2}) = -\sum_{n'L'} \frac{\left\langle 5^2 F_{5/2} nL; K \left| \vec{D}\cdot\frac{\vec{r}}{r^3} \right| 6^2 D_{3/2} n'L'; K \right\rangle^2}{\Delta E_{core} + \Delta E_{RYD}} \tag{5}$$

where

$$\Delta E_{RYD} = E(n') - E(n)$$

$$\Delta E_{CORE} = E(6^2D_{3/2}) - E(5^2F_{5/2}) = 9193 cm^{-1}$$

The adiabatic expansion depends on the assumption that of the two energy differences in the denominator of Eq. 5, the second is much smaller than the first. If this is true then the denominator can be formally expanded as:

$$\frac{1}{\Delta E_{core} + \Delta E_{RYD}} = \frac{1}{\Delta E_{core}} - \frac{\Delta E_{RYD}}{(\Delta E_{core})^2} + \frac{(\Delta E_{RYD})^2}{(\Delta E_{core})^3} - .... \qquad (6)$$

Substituting this into Eq. 5 leads to

$$E_{n,L,K}^{[2]}(6^2D_{3/2}) = -\sum_{n'L'} \frac{\left\langle 5^2F_{5/2}nL;K \left| \vec{D} \cdot \frac{\vec{r}}{r^3} \right| 6^2D_{3/2}n'L';K \right\rangle^2}{\Delta E_{core}}$$
$$+ \qquad (7)$$
$$+ \sum_{n'L'} \frac{\left\langle 5^2F_{5/2}nL;K \left| \vec{D} \cdot \frac{\vec{r}}{r^3} \right| 6^2D_{3/2}n'L';K \right\rangle^2 (\Delta E_{RYD})}{(\Delta E_{core})^2} + ...$$

The first term gives the adiabatic polarization energy due to the $6^2D_{3/2}$ state and the second gives a non-adiabatic correction. In many Rydberg systems, the first term is dominant, and expresses the net result of dipole coupling to all possible core-excited levels in a single constant. The resulting energy shift is simply evaluated for any Rydberg without the need for a specific calculation for each level.

Considering the many common factors in the two terms in Eq. 7, it can be shown that the ratio of the second term to the first is:

$$\frac{E^{[2]}_{1stNA}}{E^{[2]}_{AD}} = \frac{\sum_{n'} \left|\langle n'L'|r^{-2}|nL\rangle\right|^2 (E(n') - E(n))}{(\Delta E_{CORE})\sum_{n'} \left|\langle n'L'|r^{-2}|nL\rangle\right|^2} \qquad (8)$$

and using the properties of hydrogenic radial wavefunctions, this can be simplified to:

$$\frac{E^{[2]}_{1stNA}}{E^{[2]}_{AD}} = \frac{(4 - L(L+1) + L'(L'+1))\langle r^{-6}\rangle_{nL}}{2\langle r^{-4}\rangle_{nL}} \frac{1}{\Delta E_{CORE}(a.u.)} \qquad (9)$$

where $n$ and $L$ identify the Rydberg level in question, $L'$ is the angular momentum of the Rydberg electron in the intermediate state, and $\Delta E_{CORE}$ is the excitation energy of the core state in atomic units. Table II evaluates this ratio for several values of $L$ within the $n=27$ level of Th$^{2+}$. Similar results are found for other $n$s. It is apparent from the values shown in Table II that the adiabatic expansion fails for the portion of the second-order dipole perturbation energy coming from intermediate $6^2D_{3/2}$ core levels, at least for Rydberg levels with $L < 12$. A similar conclusion can be drawn for intermediate $6^2D_{5/2}$ core levels ($\Delta E(6^2D_{5/2})$=14486 cm$^{-1}$). Additional contributions to the second-order dipole perturbation energies from higher levels (7d, 8d, etc.) would be expected to be described accurately by the adiabatic expansion because of their higher excitation energies. Fortunately, it is not difficult to calculate explicitly the second-order dipole perturbation energies from a specific core intermediate state. For example, the contribution from intermediate states with $6^2D_{3/2}$ core state is given by

$$E^{[2]}_{n,L,K}(6^2D_{3/2}) = -\langle 5^2F_{5/2}\|\vec{D}\|6^2D_{3/2}\rangle^2 \sum_{n'L'} \begin{Bmatrix} K & L & 5/2 \\ 1 & 3/2 & L' \end{Bmatrix}^2 \frac{\left\langle nL\left\|\frac{\hat{r}}{r^2}\right\|n'L'\right\rangle^2}{\Delta E(6^2D_{3/2}) + \Delta E_{Ryd}} \qquad (10)$$

where $\Delta E(6^2D_{3/2})$ is the known excitation energy of the $6^2D_{3/2}$ level. Except for the squared dipole matrix element at the beginning of Eq. 10, everything else is known. The sum can be done numerically using Dalgarno-Lewis method [18]. In this way, the full second-order dipole perturbation energy for $6^2D_{3/2}$ intermediate core levels can be calculated up to a constant which represents the square of the reduced dipole matrix element connecting the ground $5^2F_{5/2}$ level and the $6^2D_{3/2}$ level. A similar calculation can be carried out for the second-order dipole perturbation energy due to $6^2D_{5/2}$ intermediate levels. Of course, theses calculation must be carried out for each specific $nLK$ level of interest.

This suggests a modification of the effective potential method that could successfully describe the binding energies of the $Th^{2+}$ Rydberg levels studied here. The contributions to the Rydberg binding energy due to the second-order dipole energies from $6d$ intermediate core levels are not well described by the $\alpha_{d,0}$ and $\alpha_{d,2}$ terms in Eq. 4. Therefore these energy contributions must be calculated explicitly for each $nLK$ Rydberg level, and they are known exactly except for a constant that represents the square of the dipole matrix element connecting the core ground state to each $6d$ level. The total deviation of the Rydberg binding energy from hydrogenic would then be given by the sum of these two explicitly calculated terms and the expectation value of a modified effective potential

$$\Delta E(nLK) = \left|\left\langle 5^2F_{5/2}\|D\|6^2D_{3/2}\right\rangle\right|^2 \; E^{[2]}_{n,L,K}\left(6^2D_{3/2}\right)^* + \left|\left\langle 5^2F_{5/2}\|D\|6^2D_{5/2}\right\rangle\right|^2 E^{[2]}_{n,L,K}\left(6^2D_{5/2}\right)^*$$

$$+ \left\langle (J_c)nL_K \left| V^{Mod}_{eff} \right| (J_c)nL_K \right\rangle$$

(11)

where $E^{[2]}_{n,L,K}(6^2D_{3/2})^*$ and $E^{[2]}_{n,L,K}(6^2D_{5/2})^*$ denote the second-order dipole perturbation energies calculated from these two core states assuming unit matrix elements. The modified effective potential has the same form as Eq. 4

$$V^{Mod}_{eff} = -\left[\frac{e^2 \alpha^{Mod}_{d,0}}{2r^4}\right] - \left[\left[\frac{eQ}{r^3} + \frac{\alpha^{Mod}_{d,2}}{2r^4}\right] \times \frac{X^{[2]}_c(J_c) \cdot C^{[2]}(\Omega_r)}{\begin{pmatrix} 5/2 & 2 & 5/2 \\ -5/2 & 0 & 5/2 \end{pmatrix}}\right] \quad (12)$$

with $\alpha^{Mod}_{d,0}$ and $\alpha^{Mod}_{d,2}$ representing $\alpha_{d,0}$ and $\alpha_{d,2}$ with the contributions of the *6d* levels removed. This procedure is similar to that used by Snow in treating the non-adiabatic quadrupole polarization energies in barium Rydberg levels [17].

Figure 3 illustrates the importance of the strong non-adiabatic effects on the energies of $n=27$ Rydberg levels of $Th^{2+}$. The six energy levels corresponding to $L = 9, 11,$ and 13 are shown. The left hand panel for each $L$ shows the calculated energy levels predicted in the adiabatic model, assuming theoretical values for Q, $\alpha_{d,0}$ and $\alpha_{d,2}$ [3, 19]. The right hand panel shows the energy levels computed if the *6d* levels are treated separately and their contributions are evaluated with Eq. 10 and its analog for the $6^2D_{5/2}$ level. The importance of non-adiabatic effects is clear in the $L=9$ levels, where they dramatically alter the order of levels. In the $L=13$ levels, the non-adiabatic effects are much smaller and are approximated by the first non-adiabatic corrections. This reinforces the conclusion, which could be drawn from Table II, that the adiabatic model may be adequate for very high-L levels of $Th^{2+}$, but not for levels with $L < 12$.

## C. Line Identification and Fits

Using the model described by Eq. 11, it was possible to interpret the three RESIS spectra shown in Fig. 2. Fortunately theoretical estimates [3, 19] of the five most significant core properties were available, and these formed a starting point.

$$Q = 0.62$$

$$\alpha_{d,0}^{Mod} = 8.582$$

$$\alpha_{d,2}^{Mod} = 0.054$$

$$\left|\left\langle 5^2 F_{5/2} \| \vec{D} \| 6^2 D_{3/2} \right\rangle\right| = 1.530$$

$$\left|\left\langle 5^2 F_{5/2} \| \vec{D} \| 6^2 D_{5/2} \right\rangle\right| = 0.412$$

Using these initial estimates, a simulation of each of the spectra was constructed. These initial spectra were a rather poor match to the details of Fig. 2. The high-$L$ features, the spilt and the sharpness of the left side, allowed for limits to be place on the possible values of Q and $\alpha_{d,0}^{Mod}$. Using theses limits, the five parameters were varied until a reasonable match was obtained for resolved structures in all three spectra. In spite of the complexity of the spectra and the limited experimental resolution, fourteen lines were identified as representing single resolved excitations. These fourteen lines are shown in Fig. 2 by eight different letters signifying common values of $L$ and $K$ in the lower state of the transition. Table III identifies each of these transitions and reports the difference of their transitions frequency from the appropriate hydrogenic frequency. For example, the three lines labeled "a" in Fig. 2 represent excitation of the L=10, K=8.5 levels and are visible in all three spectra at approximately the same quantum defect. Referring to Table III the transition frequency of the three "a" lines exceed the hydrogenic frequency by 3288, 3133, and 2903 MHz. The lines labeled "b" represent excitation of *L*=9, *K*=8.5 levels in *n*=27 and 28. The analogous line in *n*=29 is not well resolved and not used in the fit, but its position is indicated by the "(b)" in Fig. 2. Non-adiabatic effects are most noticeable in the excitation lines originating from *L*=7 or 8 levels. For example the excitation of *L*=8, *K*=5.5 levels, labeled "d" in Fig. 2, shifts dramatically in quantum defect in the three spectra due to non-adiabatic effects.

The frequencies of the fourteen well-resolved single lines were fit to determine the best values of the five parameters in Eq. 11. Before the fit, the transition energies were corrected to remove a small relativistic contribution

$$E_{rel}(n,L) = \frac{\alpha^2 q^4}{2n^4}\left(\frac{3}{4} - \frac{n}{L+\frac{1}{2}}\right) \quad (13)$$

and those small corrections are listed in Table III. Possible contributions due to second order quadrupole coupling to Rydberg levels bound to the $5^2F_{7/2}$ core level, at 4325cm$^{-1}$, were found to be negligible for purposes of this study [5]. The results of this fit were:

$$Q = 0.54(4)$$

$$\alpha_{d,0}^{Mod} = 9.67(15)$$

$$\alpha_{d,2}^{Mod} = 1.5(1.3)$$

$$\left|\langle 5^2F_{5/2}\|\vec{D}\|6^2D_{3/2}\rangle\right| = 1.435(10)$$

$$\left|\langle 5^2F_{5/2}\|\vec{D}\|6^2D_{5/2}\rangle\right| = 0.414(24)$$

The error in each fitted parameter is the quadrature sum of the error from the fit, expanded to reflect the quality of the fit, and the error due to uncertainty in the calibrations of the beam speed and the intersection angle. Also shown in Table III is the fitted position of each of the fourteen lines and the difference between the observed and best fit.

Additional $Th^{3+}$ properties such as the Landé factor, $g_J$, the permanent hexadecapole moment, $\Pi$, and the scalar quadrupole polarizability, $\alpha_{Q,0}$, that occur in the full effective potential [2], were included on a trial basis in the fit. These additional properties all failed to improve the fit and their fitted values were not significantly different from zero. In addition, the possible effects of Stark shifts in the upper n levels due to a constant stray electric field ≤0.05V/cm were considered and did not improve the fit. The average difference between the measured and fitted positions was on the order of 50MHz. Although this exceeds the precision of the observations by a factor of five, it is probably consistent with a combination of small errors in the beam speed or the calibration of the intersection angle, drifts of the laser frequency (≤30MHz) or beam trajectory, small stray electric fields and contributions of higher terms in $V_{eff}$.

The fitted parameters were used to estimate the positions of all excitation lines, including those unresolved in Fig. 2, and simulations were prepared of each of the observed spectra. One example is shown in Fig. 4. The simulated spectrum matches the observed spectrum quite well. All observed features are represented in the simulation, and the simulation contains no features that are not seen in the data. Some minor differences are seen, especially in regions where there are multiple overlapping lines.

**IV. DISCUSSION**

Table IV compares the properties derived here from the Rydberg spectra to theoretical estimates. The three properties that are most directly reflected in the spectra, and which are therefore most precisely determined are the quadrupole moment, Q, the portion of the scalar dipole polarizability not due to *6d* levels, $\alpha_{d,0}^{Mod}$, and the dipole matrix element connecting the ground $5^2F_{5/2}$ level to the $6^2D_{3/2}$ level. The fitted values, shown in column five of Table IV are similar to, but significantly different from the initial theoretical estimates shown in column four of Table IV. The measured value of Q, determined to about 8%, disagrees with the Dirac-Fock (DF) estimate by almost a factor of two, but is in much better agreement with the all-orders result from Relativistic Many Body Perturbation Theory (RMBPT). The dipole matrix element from $5^2F_{5/2}$ to $6^2D_{3/2}$ follows a similar pattern. The measured value is nearly a factor of two smaller than the DF estimate, but is in much better agreement with the all-orders RMBPT result. The measured result for $\alpha_{d,0}^{Mod}$ agrees to within about 10% with the RMBPT result. Almost all of this comes from the polarizability of the Rn-like $Th^{4+}$ core of $Th^{3+}$ [4]. This quantity has been independently measured in a recent experiment and agrees with a Relativistic Random Phase Approximation (RRPA) calculation to within 5% [10]. The portion of the tensor dipole polarizability not due to the *6d* levels, $\alpha_{d,2}^{Mod}$, is not very precisely determined from the spectra, but appears to be very small, in agreement with the RMBPT. In other words, the tensor polarizability is almost entirely due to coupling to the *6d* levels.

Although $\alpha_{d,0}$ and $\alpha_{d,2}$, the full adiabatic dipole polarizabilities do not directly describe the levels studied here, they can still be calculated from the fitted results. Given the fitted dipole matrix elements, the contribution of both *6d* levels to the adiabatic scalar and tensor polarizabilities can be calculated, as indicated in the appendix.

$$\alpha_{d,0} = \alpha_{d,0}^{Mod} + \alpha_{d,0}^{D_{3/2}} + \alpha_{d,0}^{D_{5/2}} = 9.67(15) + 5.46(8) + 0.29(3) = 15.42(17) \qquad (14a)$$

$$\alpha_{d,2} = \alpha_{d,2}^{Mod} + \alpha_{d,2}^{D_{3/2}} + \alpha_{d,2}^{D_{5/2}} = 1.5(1.3) - 5.46(8) + 0.33(4) = -3.6(1.3) \qquad (14b)$$

The total scalar polarizability agrees with the RMBPT result to within 2%, but this appears to be fortuitous since agreement for the major contributions is not nearly so close. The total tensor polarizability is found to be slightly smaller than the theoretical prediction.

In general, the fitted parameters confirm the predictions of the best RMBPT theoretical calculations at about the 10-20% level, but are in clear disagreement with predictions based on the DF model. This is probably not surprising since the *7s*, *5f*, and *6d* orbitals are nearly degenerate for thorium [4] indicating that correlation effects, omitted in the DF model, are likely to be especially important in thorium. In other words, even though the DF model is fully relativistic, it is a poor choice for describing this ion when even moderate precision is needed. The detailed description of the RMBPT calculations in references 3 and 4, and the references included there provide more detail about the challenges of describing such a system in an *a-priori* model.

The $Th^{3+}$ properties measured here represent a substantial addition to the experimental knowledge base for Fr-like atoms and ions. Optical spectroscopy of Fr-like ions of any charge is very limited [5]. Aside from optical spectroscopy, the only other measured properties of any Fr-like system are a few lifetimes measured in neutral Fr [20, 21, 22]. The properties reported here, therefore provide important new tests of atomic structure theory in these highly-relativistic one-valence electron systems. Although the thorium ions may present an especially difficult case because of the near degeneracy of several orbitals, this example clearly demonstrates that experimental measurements have an important role to play in clarifying the accuracy of such calculations.

Given the preliminary estimates of the Th$^{2+}$ fine structure patterns provided by the present study, it appears feasible to obtain much more precise measurements using the RESIS/microwave method [13]. Such measurements would be immune to most of the factors limiting the precision of the present study. They should also be able to fully resolve the structures studied here and explore higher $L$ levels that are completely unresolved here.

## ACKNOWLEDGMENTS

The work reported here was carried out at the J. R. Macdonald Laboratory at Kansas State University. We are thankful to the laboratory staff and management for their assistance. The work was supported by the Chemical Sciences, Geosciences, and Biosciences Division of the Office of Basic Energy Science, U.S. Department of Energy.

## APPENDIX

The scalar dipole polarizability, $\alpha_{d,0}$, and the tensor dipole polarizability, $\alpha_{d,2}$, are given by:

$$\alpha_{d,0} \equiv \frac{1}{9} \sum_{\gamma J_c'} \frac{\left\langle 5^2 F_{5/2} \| \vec{D} \| \gamma J_c' \right\rangle^2}{\Delta E(\gamma J_c')}$$

and

$$\alpha_{d,2} \equiv -\sqrt{\frac{50}{63}} \sum_{\gamma J_c'} \frac{\left\langle 5^2 F_{5/2} \| \vec{D} \| \gamma J_c' \right\rangle^2}{\Delta E(\gamma J_c')} (-1)^{J_c'-5/2} \begin{Bmatrix} J_c' & 1 & 5/2 \\ 2 & 5/2 & 1 \end{Bmatrix}.$$

where

$$\vec{D} \equiv \sum_{i=1}^{87} r_i C^{[1]}(\hat{r}_i)$$

**REFERENCES**

[1] S.R. Lundeen, in *Advances in Atomic, Molecular, and Optical Physics*, edited by C.C. Lin and P. Berman, (Academic, New Your, 2005), Vol. 52, pp.161

[2] Julie A. Keele, et. al., Phys. Rev. A **81**, 022506 (2010).

[3] U.I. Safronova, W.R. Johnson, and M.S. Safronova, Phys. Rev. A **74**, 042511 (2006)

[4] U.I. Safronova, W.R. Johnson, and M.S. Safronova, Phys. Rev. A **76**, 042504 (2007).

[5] J. Blaise and J. Wyart, *Selected Constants: Energy Levels and Atomic Spectra of Actinides*, http://www.lac.u-psud.fr/Database/Contents.html

[6] C. J. Campbell, et. al., Phys. Rev. Lett. **102**, 233004 (2009).

[7] L.R. Churchill, et. al., Phys. Rev. A **83**, 012710(2011).

[8] V. Goncharov and M. Heaven, J. Chem. Phys.**124**, 064312 (2006)

[9] M.E. Hanni, et. al., Phys. Rev. A **81**, 042512 (2010).

[10] M.E. Hanni, et. al., Phys. Rev. A **82**, 022512 (2010).

[11] R. A. Komara, et. al., J. Phys. B **38**, S87 (2005).

[12] R. F. Ward Jr., et. al., Phys. Rev. A **53**, 113 (1996).

[13] M. E. Hanni, et. al., Phys. Rev. A **78**, 062510 (2008).

[14] R. Drachman, Phys. Rev. A **26**, 1228 (1982).

[15] W. Clark, Phys. Rev. A 53, 2248, (1996).

[16] T. F. Gallagher, et. al., Phys. Rev. A **26**, 2611 (1982).

[17] E. L. Snow, et. al., Phys. Rev. A **71**, 022510 (2005)

[18] R. A. Komara et. al., Phys. Rev. A **59**, 251 (1999).

[19] U. I. Safronova (private communication)

[20] J.E. Simsarian, et. al., Phys. Rev. A **57**, 2448 (1998).

[21] J.M. Grossman, et. al., Phys. Rev. A **62**, 062502 (2000).

[22] E. Gomez, et. al., Phys. Rev. A **71**, 062504 (2005).

Figure 1. Diagram of the RESIS apparatus used in this work. An ECR source produces a 75keV beam of $Th^{3+}$, charge and mass selected at (1) using a 20° magnet. The beam then passes through a Rb target at (2) where it charge captures a highly excite electron, becoming a beam of $Th^{2+}$ Rydberg states. Then at (3) the beam passes through a 15° magnet where it is charge analyzed and the $Th^{2+}$ Rydberg beam is selected, eliminating any remaining $Th^{3+}$ beam. From there it passes through an einzel lens at (4) to remove any weakly bound Rydberg states from the beam. Rydberg transitions are then excited with a $CO_2$ laser at (5), making transitions from one $n$ level to a higher $n$ level. Then at (6) the upper $n$ Rydberg state is Stark ionized, and at (7) it is steered and focused into the channel electron multiplier.

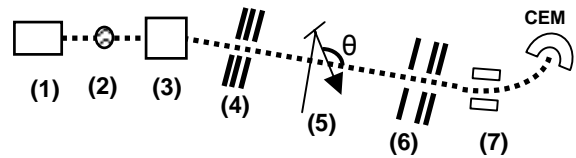

Figure 2. The RESIS spectrums of Th$^{2+}$ Rydberg states for three different transitions, each spectrum is labeled to identify the specific transition. The solid line on each of graphs is the original signal and the points with error on them are the signal times 20. On the x-axis is the quasi quantum defect discussed in the text. Each spectrum had a limited scan range. The large peak around zero is referred to as the high-*L* peak because it is made up of all the transitions with little quantum defect, the higher *L* levels that are unresolved. Letters without parentheses denote transitions used during the fitting process, while letters with parentheses identify analogous lines not used in the fit.

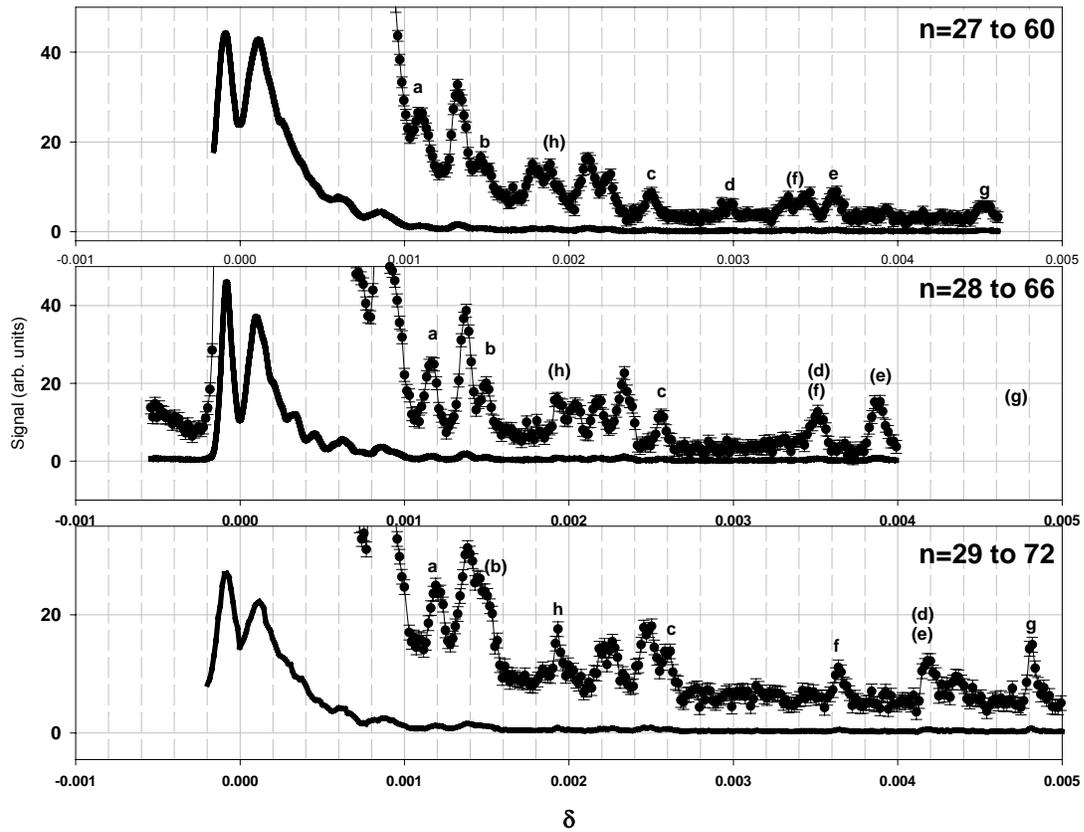

Figure 3. This figure illustrates the importance of non-adiabatic effects in the second order dipole perturbation energy. For $n=27$ $L=9$, 11, and 13 the energy levels were calculated using the adiabatic model with the theoretical values of Q, $\alpha_{d,0}$ and $\alpha_{d,2}$, these are shown on the left of each panel. On the right of each panel the non-adiabatic effects were calculated and added to the model. In both cases only the first order quadrupole energies and second order dipole energies are included. It can be observed that these non-adiabatic effects drastically affect the order of the energy levels for $L=9$, but by $L=13$ the effect is limited to slight shifts in each level, which would likely be described by the non-adiabatic terms of the effect potential.

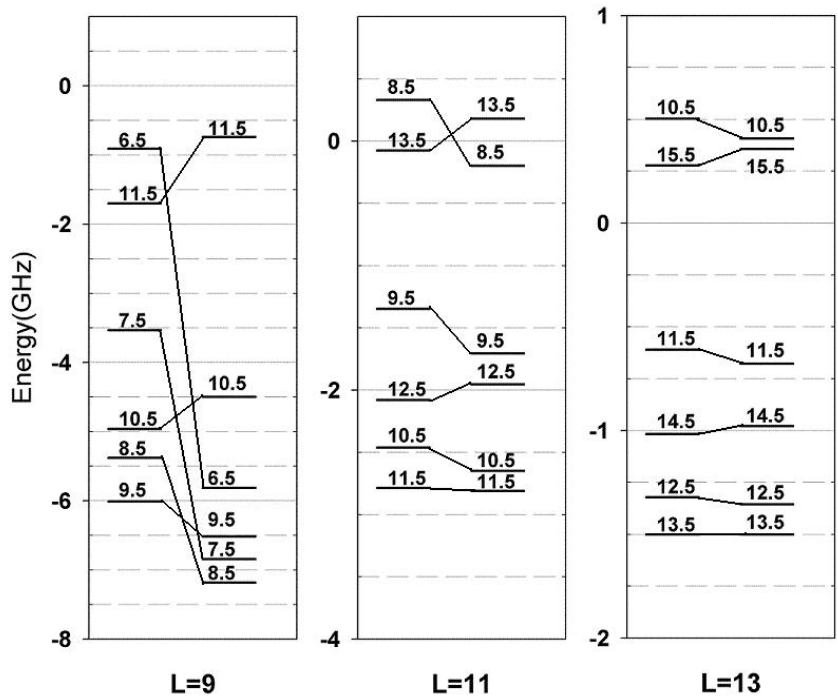

Figure 4. An example of comparison between observed and simulated spectrums using the *n*=29 to 72 spectrum. On top is the simulated spectrum using the fitted properties. On bottom is the actual experimental observation for *n*=29 to 72. The solid line in both is the signal and the line with points in both is the signal times twenty. On the x-axis is the transition's frequency minus the non-relativistic hydrogenic energy in GHz. The simulated spectrum shows good agreement with observed spectrum, with only slight variation in areas of overlapping signals.

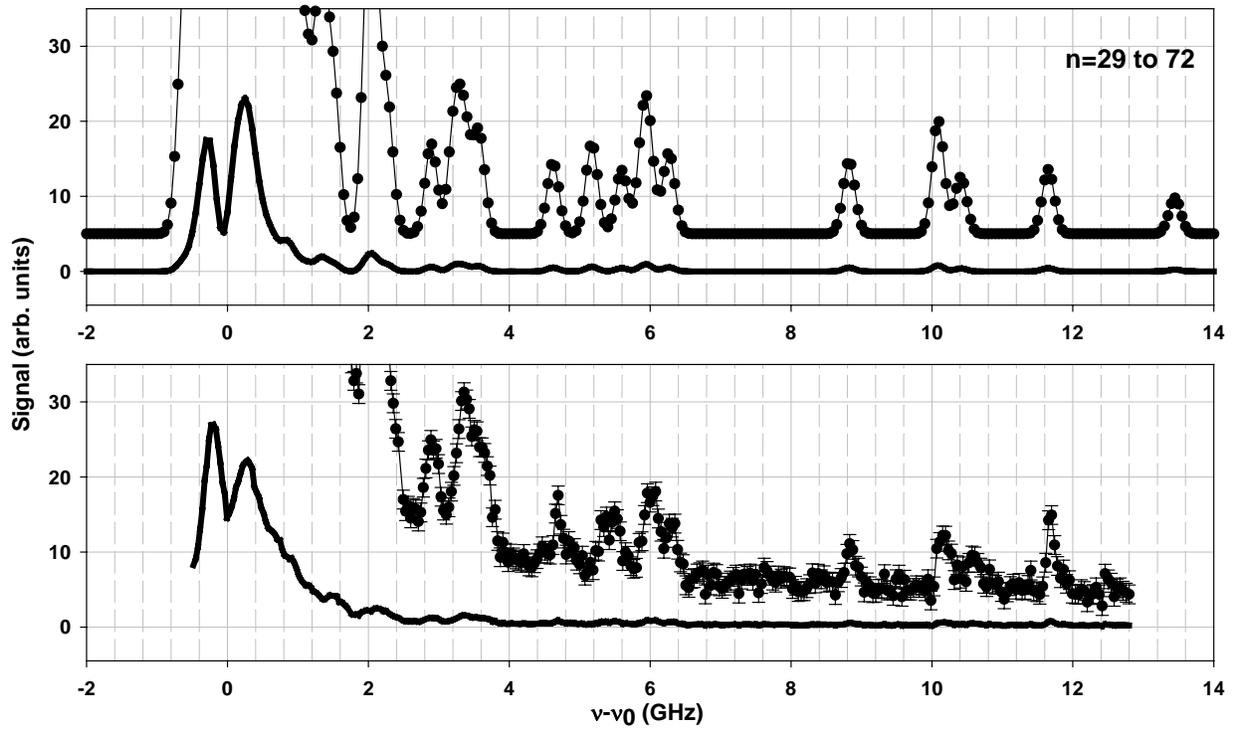

Table I. Specific RESIS transitions observed in this study are given in column one. In column two are the transition's non-relativistic hydrogenic frequencies in cm$^{-1}$. Column three shows the $CO_2$ laser line used for the transition and column four gives its frequency.

| Transitions observed $n$-$n'$ | $\nu_0$ (cm$^{-1}$) | $CO_2$ Line | $CO_2$ Laser frequency (cm$^{-1}$) |
|---|---|---|---|
| 27-60 | 1080.4358 | 9R(24) | 1081.0874 |
| 28-66 | 1033.0072 | 9P(34) | 1033.4880 |
| 29-72 | 983.8404 | 10R(34) | 984.3832 |

.

Table II. Estimate of the convergence of the adiabatic expansion of the second order dipole perturbation energy of the Th$^{2+}$ Rydberg states with $n=27$ and $6 \leq L \leq 13$ due to coupling to intermediate states in which the Th$^{3+}$ core is excited to the low-lying $6^2D_{3/2}$ level at 9193 cm$^{-1}$. The table shows the ratio of the 1st non-adiabatic term to the adiabatic term, computed from Eq. 9 of the text. Column one gives the $L$ of the $n=27$ Rydberg level in question. Columns two and three show the ratio for intermediate levels with $L' = L+1$ and $L'=L-1$ respectively. Notice the second term is actually larger than the first in the lowest $L$ levels shown, but becomes gradually smaller as $L$ increases, indicating that the adiabatic model does not describe the lowest $L$ levels but may be adequate for levels with $L \geq 12$.

| L  | Ratio($L' = L+1$) | Ratio($L' = L-1$) |
|----|-------------------|-------------------|
| 6  | 3.58              | -1.59             |
| 7  | 2.13              | -1.07             |
| 8  | 1.37              | -0.75             |
| 9  | 0.93              | -0.54             |
| 10 | 0.65              | -0.40             |
| 11 | 0.47              | -0.31             |
| 12 | 0.35              | -0.24             |
| 13 | 0.27              | -0.19             |

Table III. A summary of observed positions, corrections and fitted positions for the spectra in fig. 2. In column one the transitions are identified by (L, K) of the lower state. The transitions observed were to the upper states with L'=L+1 and K'=K+1. The second column identifies the letter the transition is denoted by on Fig 2. Column three gives the measured frequency with the error due to the fit of the peak and column four gives the relativistic correction for each transition. Then column five gives the relativistic corrected positions, $E^{[1]}$. Column six gives the fitted position of the transitions and column seven gives the difference between the fitted and corrected observed positions, from column five.

| Transition (L, K) | Fig 2. Identifier | $E^{measured}$ (MHz) | $\Delta E_{rel}$ (MHz) | $E^{[1]}$ (MHz) | Fitted positions (MHz) | Fitted–$E^{[1]}$ (MHz) |
|---|---|---|---|---|---|---|
| *n*=27 to 60 | | | | | | |
| (10, 8.5) | a | 3288(8) | 44 | 3244 | 3295 | 51 |
| (9, 10.5) | b | 4428(16) | 50 | 4378 | 4325 | -53 |
| (8, 9.5) | c | 7514(8) | 59 | 7455 | 7588 | 133 |
| (8, 5.5) | d | 8932(17) | 59 | 8873 | 8848 | -25 |
| (8, 7.5) | e | 10874(9) | 59 | 10815 | 10762 | -53 |
| (7, 8.5) | g | 13617(9) | 69 | 13548 | 13501 | -47 |
| | | | | | | |
| *n*=28 to 66 | | | | | | |
| (10, 8.5) | a | 3133(8) | 41 | 3092 | 3077 | -15 |
| (9, 10.5) | b | 4030(8) | 47 | 3983 | 3924 | -59 |
| (8, 9.5) | c | 6905(17) | 54 | 6851 | 6791 | -60 |
| | | | | | | |
| *n*=29 to 72 | | | | | | |
| (10, 8.5) | a | 2903(7) | 38 | 2865 | 2891 | 26 |
| (8, 9.5) | c | 6320(16) | 50 | 6270 | 6262 | -8 |
| (8, 8.5) | f | 8851(8) | 50 | 8801 | 8815 | 14 |
| (7, 8.5) | g | 11691(8) | 59 | 11632 | 11643 | 11 |
| (7, 9.5) | h | 4695(8) | 59 | 4636 | 4634 | -2 |

Table IV. Comparison between $Th^{3+}$ properties derived from $Th^{2+}$ Rydberg spectra and calculated properties. Theoretical predictions are from reference [3] and include uncorrelated results from Dirac-Fock theory in column two, and results from two levels of approximation of relativistic many-body perturbation theory (RMBPT) in columns three and four. Column three represents RMBPT result complete through third-order. Column four represents results from all-orders RMBPT including single and double excitations. Numerical results for Q were obtained in a private communication [19]. Column five gives the result from this study. All are in atomic units.

| Property | Theory(DF) | Theory(DF+2+3) | Theory(SD) | Expt. |
|---|---|---|---|---|
| Q | 0.91 | ------ | 0.62 | 0.54(4) |
| $\alpha_{d,0}$ | ---- | 13.523 | 15.073 | 15.42(17) |
| $\alpha_{d,0}^{Mod}$ | ---- | 8.562 | 8.582 | 9.67(15) |
| $\alpha_{d,2}$ | ---- | -4.763 | -6.166 | -3.6(1.3) |
| $\alpha_{d,2}^{Mod}$ | ---- | -0.018 | 0.054 | 1.5(1.3) |
| $|<5^2F_{5/2}||D||6^2D_{3/2}>|$ | 2.428 | 1.337 | 1.530 | 1.435(10) |
| $|<5^2F_{5/2}||D||6^2D_{5/2}>|$ | 0.639 | 0.362 | 0.412 | 0.414(24) |